%% file: main.tex
\documentclass[10pt,conference]{IEEEtran}
\IEEEoverridecommandlockouts
\input{macros2}
\begin{document}

\title{Just-in-Time Detection of Silent Security Patches}

\author{\IEEEauthorblockN{1\textsuperscript{st} Xunzhu Tang}
\IEEEauthorblockA{\textit{University of Luxembourg} \\
Luxembourg \\
xunzhu.tang@uni.lu}
\and
\IEEEauthorblockN{2\textsuperscript{nd} Kisub Kim}
\IEEEauthorblockA{\textit{Singapore Management University} \\
Singapore\\
falconlk00@gmail.com}
\and 
\IEEEauthorblockN{3\textsuperscript{rd} Saad Ezzini }
\IEEEauthorblockA{\textit{Lancaster University} \\
United Kindom\\
s.ezzini@lancaster.ac.uk}
\and
\IEEEauthorblockN{4\textsuperscript{th} Yewei Song }
\IEEEauthorblockA{\textit{University of Luxembourg} \\
Luxembourg \\
yewei.song@uni.lu}
\and
\IEEEauthorblockN{5\textsuperscript{th} Haoye Tian }
\IEEEauthorblockA{\textit{University of Luxembourg} \\
Luxembourg \\
tianhaoyemail@gmail.com}
\and
\IEEEauthorblockN{6\textsuperscript{th} Jacques Klein }
\IEEEauthorblockA{\textit{University of Luxembourg} \\
Luxembourg \\
jacques.klein@uni.lu}
\and
\IEEEauthorblockN{7\textsuperscript{th} Tegawendé F. Bissyandé }
\IEEEauthorblockA{\textit{University of Luxembourg} \\
Luxembourg \\
tegawende.bissyande@uni.lu}
}

\maketitle

\input{sections/0.abstract}

\input{sections/1.introduction}
\input{sections/3.approach}
\input{sections/4.experiemtal_design}
\input{sections/5.experiment_result}

\input{sections/8.threats}

\input{sections/2.relatedwork}

\input{sections/6.conclusion}

\bibliographystyle{IEEEtran}
\balance
\bibliography{references}

\end{document}

%% file: macros2.tex
\def\BibTeX{{\rm B\kern-.05em{\sc i\kern-.025em b}\kern-.08em
    T\kern-.1667em\lower.7ex\hbox{E}\kern-.125emX}}

\usepackage{adjustbox}
\usepackage{algorithm}
\usepackage{algorithmicx}
\usepackage{algpseudocode}
\usepackage{array}
\usepackage{balance}
\usepackage{booktabs}
\usepackage[skip=1pt,labelfont=bf]{caption}
\usepackage{calc}
\usepackage{calligra}
\usepackage{color}
\usepackage{colortbl}
\usepackage{courier}
\usepackage{csvsimple}
\usepackage{enumitem}
\usepackage{fancybox}
\usepackage{fontenc}
\usepackage{fontawesome5}
\usepackage{graphicx}
\usepackage{listings}
\usepackage{longtable}
\usepackage{lscape}
\usepackage{makecell}
\usepackage{marvosym}
\usepackage{moreverb}
\usepackage{multicol}
\usepackage{multirow}
\usepackage{pifont}
\usepackage{pgfplots}
\usepackage{rotating}
\usepackage{setspace}
\usepackage{siunitx}
\usepackage{soul}
\usepackage{subcaption}
\usepackage{tablefootnote}
\usepackage[most]{tcolorbox}
\usepackage{threeparttable}
\usepackage{tikz}
\usepackage[normalem]{ulem}
\usepackage{url}
\usepackage{wasysym}
\usepackage{xspace}

\usepgfplotslibrary{statistics}
\usetikzlibrary{matrix, shapes.geometric, arrows}
\usetikzlibrary{shapes, arrows, positioning}

\algnewcommand\algorithmicforeach{\textbf{for each}}
\algdef{S}[FOR]{ForEach}[1]{\algorithmicforeach\ #1\ \algorithmicdo}

\newcolumntype{L}[1]{>{\raggedright\let\newline\\\arraybackslash\hspace{0pt}}m{#1}}
\newcolumntype{C}[1]{>{\centering\let\newline\\\arraybackslash\hspace{0pt}}m{#1}}
\newcolumntype{R}[1]{>{\raggedleft\let\newline\\\arraybackslash\hspace{0pt}}m{#1}}

\definecolor{codegreen}{rgb}{0,0.6,0}
\definecolor{codered}{rgb}{1,0,0}
\definecolor{codegray}{rgb}{0.5,0.5,0.5}
\definecolor{codepurple}{rgb}{0.58,0,0.82}
\definecolor{backcolour}{rgb}{0.95,0.95,0.92}
\definecolor{lightgray}{gray}{0.9}

\newboolean{showcomments}
\setboolean{showcomments}{true}
\ifthenelse{\boolean{showcomments}}
 { \newcommand{\mynote}[2]{
      \fbox{\bfseries\sffamily\scriptsize#1}
        {\small$\blacktriangleright$\textsf{\emph{#2}}$\blacktriangleleft$}}}
        { \newcommand{\mynote}[2]{}}
        
\definecolor{DarkOrange}{rgb}{0.8,0.3,0.0}
\definecolor{DarkCyan}{rgb}{0.0, 0.55, 0.55}
\definecolor{DarkCyel}{rgb}{1.0, 0.49, 0.0}
\definecolor{yellow-green}{rgb}{0.6, 0.8, 0.2}

\newcolumntype{?}{!{\vrule width 1pt}}

\newcommand{\toolname}{\emph{\sc llmda}\xspace}

\newcommand{\find}[1]{
\begin{tcolorbox}[leftrule=0.4mm,rightrule=0mm,toprule=0mm,bottomrule=0mm,left=0.0pt,right=0.0pt,top=0pt,bottom=0pt]
\em #1
\end{tcolorbox}
}

\lstdefinelanguage{mymarkdown}{
    morekeywords={*,\#, \#\#, \#\#\#},
    sensitive=false,
    morecomment=[l]{//},
    morestring=[b]",
    commentstyle=\color{codegreen},
    keywordstyle=\color{magenta},
    numberstyle=\tiny\color{codegray},
    stringstyle=\color{codepurple},
    basicstyle=\small,
    breakatwhitespace=false,         
    breaklines=true,
    breakindent=0pt,
    keepspaces=true,                 
    numbers=left,                    
    numbersep=5pt,                  
    showspaces=false,                
    showstringspaces=false,
    showtabs=false,                  
    tabsize=2,
}

\lstdefinestyle{mystyle}{
    commentstyle=\color{codegreen},
    keywordstyle=\color{magenta},
    numberstyle=\small\color{black},
    stringstyle=\color{codepurple},
    basicstyle=\scriptsize\ttfamily,
    breakatwhitespace=false,
    breaklines=true,
    captionpos=b,
    keepspaces=true,
    showspaces=false,
    showstringspaces=false,
    showtabs=false,
    tabsize=2
}

\lstdefinelanguage{diff}{
  morecomment=[f][\color{blue}]{@@},     
  morecomment=[f][\color{red}]-,         
  morecomment=[f][\color{codegreen}]+,       
  morecomment=[f][\color{red}]{---}, 
  morecomment=[f][\color{codegreen}]{+++},
  numberstyle=\tiny\color{codegray},
  numbers=left,                    
  numbersep=5pt,         
}

\setlist{noitemsep} 

\definecolor{darkpastelred}{rgb}{0.76, 0.23, 0.13}
\definecolor{ao(english)}{rgb}{0.0, 0.5, 0.0}

\definecolor{darkpastelred}{rgb}{0.76, 0.23, 0.13}
\definecolor{ao(english)}{rgb}{0.0, 0.5, 0.0}

\newboolean{useblue}
\setboolean{useblue}{false} 

\newcommand{\maybeblue}[1]{%
    \ifthenelse{\boolean{useblue}}%
    {\textcolor{blue}{#1}}%
    {#1}%
}

%% file: sections/0.abstract.tex
\begin{abstract}
Open-source code is pervasive. In this setting, embedded vulnerabilities are spreading to downstream software at an alarming rate. While such vulnerabilities are generally identified and addressed rapidly, inconsistent maintenance policies may lead security patches to go unnoticed. Indeed, security patches can be {\em silent}, i.e., they do not always come with comprehensive advisories such as CVEs. This lack of transparency leaves users oblivious to available security updates, providing ample opportunity for attackers to exploit unpatched vulnerabilities. Consequently, identifying silent security patches just in time when they are released is essential for preventing n-day attacks, and for ensuring robust and secure maintenance practices. With \toolname we propose to (1) leverage large language models (LLMs) to augment patch information with generated code change explanations, (2) design a representation learning approach that explores code-text alignment methodologies for feature combination, (3) implement a label-wise training with labelled instructions for guiding the embedding based on security relevance, and (4) rely on a probabilistic batch contrastive learning mechanism for building a high-precision identifier of security patches. We evaluate \toolname on the PatchDB and SPI-DB literature datasets and show that our approach substantially improves over the state-of-the-art, notably GraphSPD by 20\% in terms of F-Measure on the SPI-DB benchmark.  

\end{abstract}

%% file: sections/1.introduction.tex
\section{Introduction}\label{sec:intro}
According to a recent market report\footnote{\url{https://gitnux.org/open-source-software-statistics/}}, 96\% of applications have at least one open-source component, while open-source code makes about 80\% of a given modern application. These impressive statistics indicate that open-source software (OSS) is a key element whose engineering should be closely monitored: vulnerabilities in OSS will spread to a broad range of downstream software systems. Once discovered, they enable attackers to perform ``n-day'' attacks against unpatched software systems. 

Timely software patching remains the first defense against attacks exploiting OSS vulnerabilities~\cite{li2017large, tan2021locating}. Unfortunately, security patches can go unnoticed. On the one hand, the ever-increasing number of submitted patches and security advisories can overwhelm reviewers and system administrators. 
On the other hand, the complexity of patch management processes and the inconsistency of OSS maintenance policies can lead to the release of \textit{silent security patches}. 
Such patches are submitted to the OSS repository but no specific notice is provided for maintainers of downstream software systems. Silent security patches lead to unfortunate delays in software updates~\cite{dissanayake2022and}.

Detecting silent security patches is a timely research challenge that has gained traction in the literature. 
Overall, the variety of proposed approaches attempt to analyze code changes and commit logs within patches in order to derive security relevance. However, on the one hand, the semantics of code changes are challenging to precisely extract statically. Patches are further often non-atomic, meaning that beyond a security-relevant code change, other cosmetic or non-security changes are often involved. On the other hand, commit messages, which are supposed to  describe precisely the intention of the code changes, are often missing, mostly lacking sufficient information, and sometimes misleading.  

Recent literature has largely seen machine learning as an opportunity for improving the performance of detection systems. In general, the proposed approaches~\cite{wang2020machine,wang2019detecting,tian2012identifying} build on syntactic features. Some other approaches~\cite{zhou2021spi,wang2021patchrnn} have explored deep neural networks by considering patches as sequential data. However, most recently, Wang {\em et al.}~\cite{wang2023graphspd}  have claimed that all the aforementioned methods actually ignore the program semantics and are therefore facing a high rate of false positives. They developed GraphSPD, the incumbent state-of-the-art approach in security patch detection, which models semantics based on the graph structure of the source code. Nevertheless, while the novel technique proposed in GraphSPD~\cite{wang2023graphspd} successfully captures context within patches and largely outperforms other existing techniques, it is worth noting that it focuses on local code segments, which does not allow to capture the broader context of how functions or modules interact. 





 To cope with the aforementioned challenges, our intuition is threefold: \ding{182} First, the security relevance of a patch could be better identified if a proper and detailed {\em explanation of code changes} can be obtained. To that end, we look towards the current wave of Large Language Models (LLMs), where various studies~\cite{li2023starcoder, sun2023automatic, su2023semantic}  have demonstrated their capabilities in effectively capturing the essential context and tokens within source code for a variety of tasks.  \ding{183} Second, the patch representation must effectively learn to {\em combine and align features} from the code changes with features of the change descriptions to maximally capture the relevant details for security relevance identification. \ding{184} Third, a {\em language-centric approach} where natural language \textit{instructions} are used within the inputs to guide the learning could help exploit the power of existing general models as shown in recent papers for various tasks~\cite{wei2021finetuned, chung2022scaling, dai2023instructblip}.
 
{\bf This paper.} We design and implement \toolname (read $\lambda$), an effective learning-based approach for detecting security patches. \toolname takes multi-modal inputs that it aligns into a single comprehensive representation of patches suitable for the task of security detection. The main input is the set of code changes within a patch. If available, a developer-provided description (commit log) is considered. 
\toolname further includes LLM-generated explanation of code changes in a data augmentation strategy. Inspired by prior works~\cite{wei2021finetuned, chung2022scaling}, we also adopt an instruction-finetuning methodology to better steer the model towards accounting for the specificities of the target task. Finally, once the patch embeddings are generated, \toolname designs a stochastic contrastive learning model~\cite{oh2018modeling} for predicting whether a patch is security relevant or not.

\toolname implementation is based on CodeT5+~\cite{wang2023codet5+}, and LLaMa-7b~\cite{touvron2023llama} for generating embeddings for code and text input modalities respectively. Given that these models produce different embedding spaces, we propose a new approach, named {\em PT-Former}, to align and concatenate the different embeddings. PT-Former thus takes multi-modal inputs and deploys self-attention, cross-attention, and feedforward modules to yield a single embedding. Embeddings of different patches (and their associated descriptions and generated explanations) are then grouped into batches for contrastively learning to identify security patches.  

Our contributions are as follows:
\begin{itemize}[leftmargin=*]
    \item We introduce \toolname as a novel framework for security patch detection. \toolname can detect silent security patches as it does not require any explicit descriptive information from developers to operate. It leverages LLMs for both data augmentation (generation of explanations) and patch analysis (generation of representations). It further deploys a specialized PT-Former module to align various modalities within a single embedding space, enabling the approach to extract richer information from the joint context of code and descriptions. Leveraging contrastive learning on the yielded embeddings, \toolname is able to precisely identify security patches. 
    \item We achieve new state-of-the-art performance in security patch detection. The experimental results show that our language-centric approach consistently outperforms the baseline methods (i.e., TwinRNN~\cite{wang2021patchrnn} and GraphSPD~\cite{wang2023graphspd}) on two target datasets (i.e., PatchDB~\cite{wang2021patchdb} and SPI-DB~\cite{zhou2021spi}): \toolname achieves up to $\sim$42\% and $\sim$20\% performance improvement over the incumbent state-of-the-art on both datasets, respectively.
    \item We experimentally demonstrate through ablation studies that the different components and key design decisions of \toolname are contributing to its overall performance. Notably, we show that the representations have a high discriminative power and that the yielded classification model is relatively robust (compared to the incumbent state-of-the-art).
\end{itemize}

\begin{figure*}[t!]
    \centering
    \includegraphics[width=\linewidth]{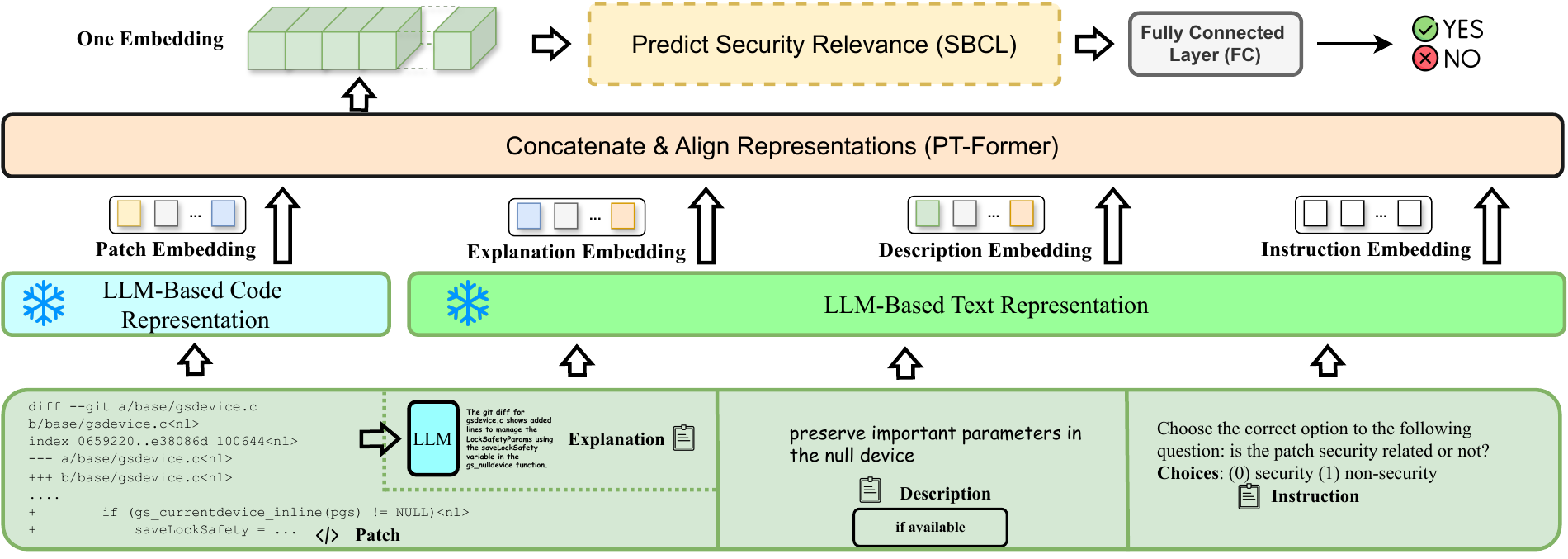}
    \caption{Overview of \toolname}
    \label{fig:overview}
    \vspace{-0.5cm}
\end{figure*}

%% file: sections/3.approach.tex

\section{The \toolname approach}
Figure~\ref{fig:overview} depicts the overview of the different steps of \toolname. First, representations of multi-modal inputs (code and texts) are obtained using LLMs. Then, the obtained representations are aligned within a unique embedding space and fused into a single comprehensive representation by the {\bf PT-Former} module. Finally, a stochastic batch contrastive learning ({\bf SBCL}) mechanism is deployed to make the predictions of whether a given patch is a security patch or not.


\subsection{Data augmentation with LLMs} \label{data}
The intention behind code changes is supposed to be provided in the patch description. Such information is then expected to be essential for security patch detection. Unfortunately, commit messages, which are meant to convey patch descriptions, are often missing, mostly non-sufficiently detailed, and even sometimes misleading. In \toolname, we explore the power of LLMs, which have demonstrated remarkable capabilities on a broad spectrum of tasks~\cite{tian2023chatgpt}, in explaining patches. As illustrated in Figure~\ref{fig:overview}, each patch is used to prompt ChatGPT (version 3.5), to produce a natural language explanation based on the following prompt instruction: ``{\em Could you provide a concise summary of the specified patch?}''\footnote{We have experimented with a variety of variations for this prompt and obtained similar outputs.}. 

Beyond the augmentation of input data with generated explanations, we also consider augmenting the representation. In transformer-based models, a typical [CLS] token is used to represent the classification token. 
It is generally positioned at the beginning of the input
sequence, serving as a signal for the model to generate a representation
suitable for classification tasks. In \toolname, we propose to specialize the classification task through a label-wise training process. The embedding of a specialized instruction for security classification is therefore added to accompany every patch input. The instruction is as follows: ``{\em Choose the correct option to the following question: is the patch security related or not? Choices: (0) security (1) non-security}''.


\subsection{Generation of bimodal input embeddings}
\toolname operates with bimodal inputs: code in the form of program patches, and text in the form of natural language description of code changes as well as the instruction for label-wise training.
We generate embeddings for each input using an adapted deep representation learning model. 

\noindent
\textbf{Patch Embeddings}: We build on CodeT5+ to infer the representation of patches.
This pre-trained model is known to be one the best-performing\footnote{\url{https://huggingface.co/Salesforce/codet5-small}} models for code representation learning. 
Given a code snippet, which is a sequence of tokens {\it C} =\{c$_{1}$, c$_2$, $\dots$, $c_n$\}, 
\textit{P} $\in \mathcal{R}^{n\times dim}$ is the associated matrix representation where each row corresponds to the representation of a token in $C$, and $dim$ is the dimension of the token embeddings.
We then employ the transformation function $f_{\text{CodeT5+}}$ on \textit{P} to yield the patch embedding $E_{p}$:

\begin{equation}
E_{p} = f_{\text{CodeT5+}}(P) = \mathcal{F}(P \cdot W_{p} + b_{p})
\end{equation}
where \( W_{p} \) is a weight matrix, \( b_{p} \) is a bias vector, and \( \mathcal{F} \) denotes a non-linear activation function. 

\noindent
\textbf{Text Embeddings}:
We leverage LLaMa-7b for the representation of text input. 
This pre-trained LLM stands out in the literature for its robust generalization capabilities across diverse domains without the need for extensive fine-tuning.
Similarly to the embedding process for patches, for a sequence of textual tokens, we build its matrix representation \textit{T} $\in \mathcal{R}^{m \times dim}$
using the initial embedding layer of a neural network model, where \textit{m} is the length of the sequence.
We then employ the transformation function \( f_{\text{LLaMa}} \) on \textit{T} and then produce a text embedding \( E_{t} \):

\begin{equation}
E_{t} = f_{\text{LLaMa}}(T) = \mathcal{G}(T \cdot W_{t} + b_{t})
\end{equation}
where \( W_{t} \) is a weight matrix for the textual transformation, \( b_{t} \) is the corresponding bias vector, and \( \mathcal{G} \) is a non-linear activation function.

\toolname is fed with three text inputs: generated code change explanations, developer-provided patch descriptions, and the instruction. 
Using the aforementioned process, we produce embeddings  $E_{t}^{ex}$, $E_{t}^{desc}$ and $E_{t}^{inst}$ respectively for each input.




\subsection{PT-Former: Embeddings alignment and Concatenation}
As that the given two embeddings $E_{p}$ and $E_{t}$ represent two different modalities, a patch and a text, their feature spaces differ. In order to leverage pre-trained unimodal models for silent security patch detection, it is key to facilitate cross-modal alignment. In this regard, existing methods (e.g. BLIP2~\cite{li2023blip}, InstructBLIP~\cite{dai2023instructblip}) resort to an image-text alignment, which we show is insufficient to bridge the modality gap. There is thus a need to align the embedding spaces before concatenating the relevant embeddings to produce a comprehensive representation of the input for the training of the classification model.

\begin{figure}[!t]
    \centering
    \includegraphics[width=0.48\textwidth]{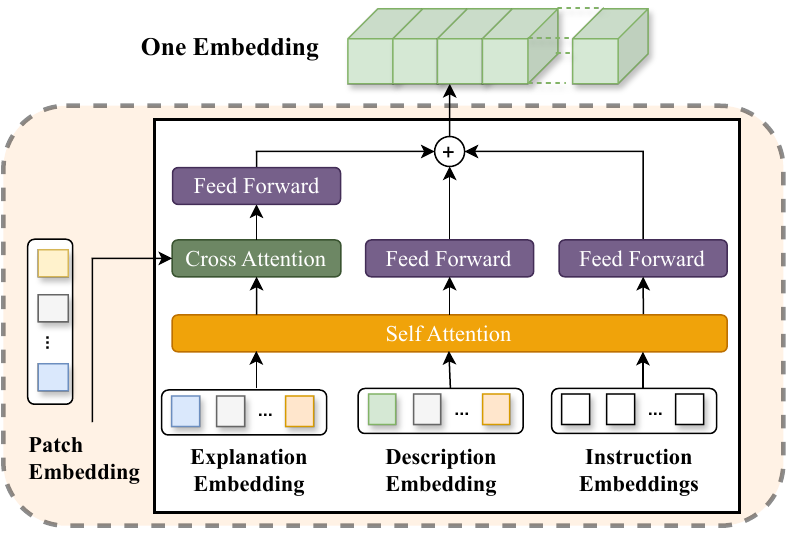}
    \caption{Architecture of PT-Former.}
    \label{fig:ptformer}
    \vspace{-0.5cm}
  \end{figure}

Figure~\ref{fig:ptformer} overviews {\em PT-Former}, a new architecture that we have designed for aligning embedding spaces and fusing the embeddings of \toolname's bi-modal inputs.
 With \textit{PT-Former}, we employ a self-attention mechanism to update all embeddings for a generated explanation, the human patch description, and the devised instruction. 
 We leverage a cross-attention module between the patch embedding and the updated explanation module. 
 Feed-forward layers are then used to align the matrix size of all hidden states before concatenating all three embeddings into a single output embedding. 
 
\noindent
\textbf{Self-Attention Mechanism (SA).}
\noindent
The self-attention mechanism is a fundamental component of the transformer architecture, designed to model interactions between elements in a sequence, 
enhancing the representation of each element by aggregating information from all other elements~\cite{devlin2018bert}. 
Because attention allows for a dynamic weighting of the importance of inputs' contribution to the representation of others, exploiting it in \toolname will enable it to understand contextual relationships within the input data.
In PT-Former, we implement a multi-head attention mechanism with \textit{h} heads to capture various aspects of these interactions, initializing each head's query (\textit{Q}), key (\textit{K}), and value (\textit{V}) matrices with values drawn from a standard normal distribution:
\begin{equation}
W_{Q_i}, W_{K_i}, W_{V_i} \sim \mathcal{N}(0, 1), \quad i = 1,...,h
\label{eq:qkv}
\end{equation}
where \textit{Q}, \textit{K}, and \textit{V} are respectively the query, key, and value for each embedding to be calculated inside the self-attention.

Consider for example the weight matric of the explanation metric $E_{t}^{ex}$.
Our self-attention mechanism over $E_{t}^{ex}$ (simply noted $E_{ex}$) is computed as:

\begin{equation}
   \footnotesize
\hat{E}_{ex} = \mathcal{SA}(E_{ex}) = \text{Softmax}\left( \frac{E_{ex} W_{Q_i} (E_{ex} W_{K_i})^T}{\sqrt{dim}} \right) E_{ex} W_{V_i}
\end{equation}
where \textit{dim} represents the dimensionality of the embeddings. 
Similarly, the two other text embeddings (i.e., $E_{t}^{desc}$ and $E_{t}^{inst}$) are passed through the {\scriptsize $\mathcal{SA}$} operation to obtain their updated embeddingds,  we will obtain updated embeddings, $\hat{E}_{t}^{desc}$ and $\hat{E}_{t}^{inst}$ respectively.

\noindent
\textbf{Cross-Attention for Alignment (CA).}
Cross-attention mechanisms have proven to be very effective in linking the semantic spaces between different types of data~\cite{wang2023codet5+}\cite{wei2020multi}.
We employ CA to align the embedding spaces of code changes ($E_{p}$), yielded by CodeT5+, and explanations ($E_{t}^{ex}$), yielded by LLaMa-7b. We focus on \textit{explanation}, since it is the main text input that we associate to the patch: description can be missing while instruction is always the same. It is however noteworthy that all text inputs are embedded with LLaMa-7b and are thus in the same embedding space as explanation.
The key feature of cross-attention is its ability to selectively focus on and integrate relevant information from both code and natural language explanations. This helps in achieving a better understanding of the relationship between the syntactical structure of code and its interpretation in natural language. 
The cross-attention computation therefore explicitates the interaction between code changes ($E_{p}$) and their explanations ($E_{t}^{ex}$).
CA starts by transforming $E_{p}$ and $\hat{E}_{t}^{expl}$ into query ($Q_{pa}$), key ($K_{ex}$), and value ($V_{ex}$) matrices using learnable weights. The attention mechanism then calculates how much focus each part of the code changes should give to different parts of the explanations. This is done by computing attention scores, which determine the output, effectively linking code changes to their explanations. The process is summarized as follows:
\begin{equation}
    \footnotesize
\begin{split}
Q_{pa} = E_{pa}W^Q, & \quad K_{ex} = E_{ex}W^K, \quad V_{ex} = E_{ex}W^V, \\
E_{pa-ex} &= \text{softmax}\left(\frac{Q_{pa}K_{ex}^T}{\sqrt{dim}}\right)AV_{ex}
\end{split}
\end{equation}
where $E_{ex}=\hat{E}_{t}^{expl}$ the updated embedding of explanation input through Self-Attention, $W^Q$, $W^K$, and $W^V$ are the weight matrices to be learned. $E_{pa-ex}$ is the fused embedding of $E_{pa}$ and $E_{ex}$.


\noindent\textbf{Embedding Fusion and Non-linear Transformation.}
We then pass the updated embeddings to feedforward layers. 
Each feedforward process involves two dense layers with a ReLU activation. 
We represent the feedforward process by the function $FF(\dots)$. 
Then, ${E}_{pa-ex}$, $\hat{E}_{desc}$, $\hat{E}_{inst}$ can be updated as $E_{pa-ex}$~=~$FF(E_{pa-ex})$, $E_{desc}$~=~$FF(\hat{E}_{desc})$, and $E_{inst}$~=~$FF(\hat{E}_{inst})$.

After obtaining attention outputs from all heads, we concatenate them to generate one embedding:
\begin{equation}
E = E_{pa-ex} \oplus E_{desc} \oplus E_{inst}
\label{eq:conca}
\end{equation}
where $\oplus$ is the concatenation operation.

\noindent 
\textbf{Label-wise Attention with Instruction.}
Inspired by the the results of InstructionBLIP~\cite{dai2023instructblip}, we postulate that an instruction
that combines a question with explicit labels can provide two advantages in our security detection task: 
first, it can provide guidance to train models in the direction of answering the security question; 
second, since it can provide the opportunity to build a relationship between inputs and the instruction labels through the calculation of their high-dimensional embeddings, leading the model to leverage instructions in a label-wise manner. 
In conclusion, the design of the instruction and its embedding within will help guide the model to focus on particular aspects of the data, thereby improving the representational efficiency for our targeted downstream task.

\noindent
\subsection{Stochastic Batch Contrastive Learning (SBCL)} 
Once {\em PT-Former} outputs a single embedding for each sample to be assessed, we must learn to predict whether it is a security patch or not. At this point, the patch is represented along with its LLM-generated explanation, developer description as well as the labelled instruction in {\em PT-Former}.
\toolname must therefore feed it into a binary classifier for predicting security relevance (cf. Figure~\ref{fig:overview}).

To enhance the learning process by effectively leveraging the intrinsic patterns within the dataset, we design a Stochastic Batch Contrastive Learning (SBCL) mechanism for security patch identification.
{\em SBCL} is designed to operate on batches of data comprising fused embeddings of security-related and non-security-related inputs (i.e., \textit{E} in Eq.~\ref{eq:conca})

Given a batch of data $\mathcal{B}$ containing embeddings $E = \{E_{pa-ex}, E_{de}, E_{in}\}$ for each data point, we employ a stochastic batch contrastive learning mechanism to discern between security and non-security data points. For each batch, we randomly select an anchor data point related to security. We then identify positive samples within the batch that are also related to security and negative samples that are not. This forms a triplet for each anchor comprising the anchor, positive, and negative samples.

\begin{figure*}
    \centering
    \includegraphics[width=\linewidth]{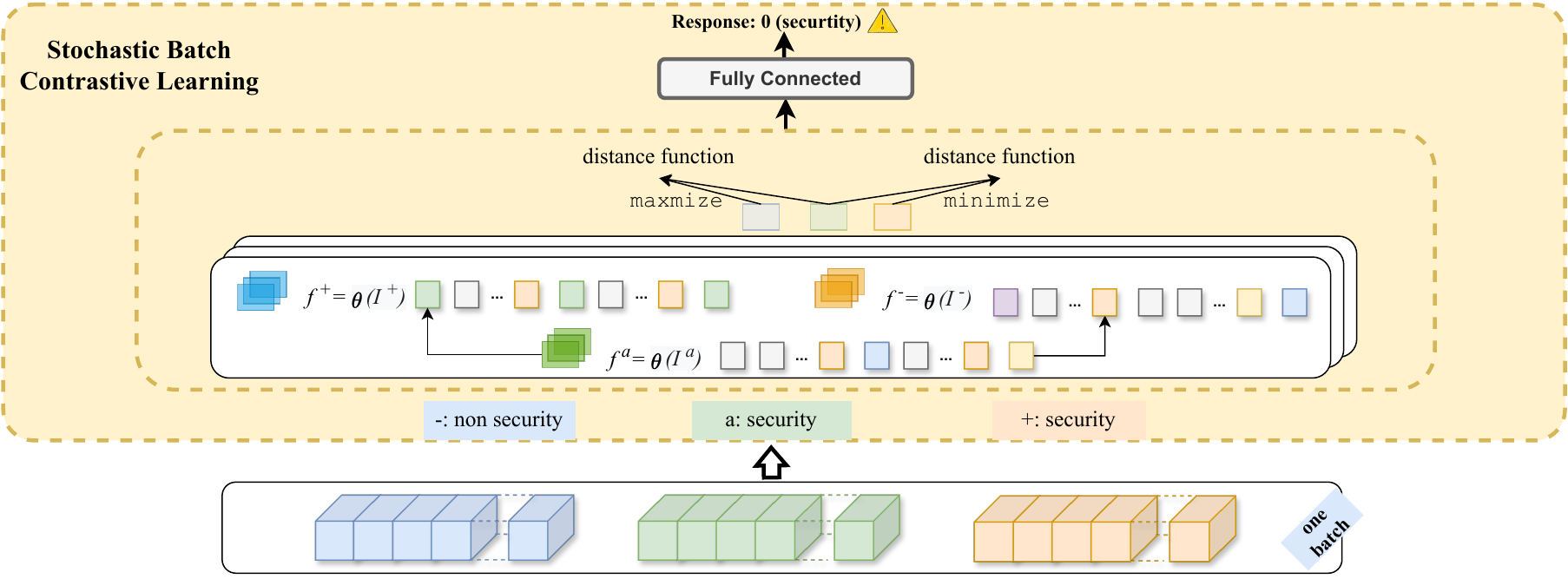}
    \caption{Overview of our SBCL layer.}
    \label{fig:sbcl}
    \vspace{-0.2cm}
\end{figure*}

\noindent
\textbf{Batch Sampling and Triplet Formation.} 
In the context of SBCL, each batch \(\mathcal{B}\) is carefully constructed to include a balanced mix of security-related (\textit{security}) and non-security-related (\textit{non-security}) examples. From each batch, we systematically form triplets for training. A triplet consists of an anchor (\(a\)), a positive example (\(p\)), and a negative example (\(n\)). The anchor and positive examples are drawn from the \textit{security} category, ensuring they share underlying security-relevant features, whereas the negative example is selected from the \textit{non-security} category.



\noindent
\textbf{Batch Mining of Positive and Negative Pairs.} 
In the SBCL framework, a systematic approach is employed to select positive and negative pairs within each batch.
 This process utilizes embeddings generated by {\em PT-Former} for all examples in a batch. The selection criterion for a positive example is its reduced similarity to the anchor, aimed at maximizing intra-class variability. Conversely, a negative example is deemed challenging based on its increased similarity to the anchor, designed to augment the model's precision in distinguishing between closely associated examples of different classes.

The selection of informative positive and negative pairs is facilitated by measuring the distances between embeddings in the batch. The Euclidean distance formula is applied to determine the distance \(d(E_a, E_b)\) between two embeddings \(E_a\) and \(E_b\):
\begin{equation}
d(E_a, E_b) = \sqrt{\sum_{i=1}^{dim} (E_a^{(i)} - E_b^{(i)})^2}
\end{equation}
This methodological approach ensures the identification and utilization of the most relevant examples for enhancing the discriminative capability of the model.

\noindent
\textbf{Stochastic Batch Contrastive Loss.} 
We design the stochastic batch contrastive loss to optimize the embedding space in order to distinguish between security-related and non-security-related examples effectively. This objective is achieved by minimizing the distance between embeddings of anchor and positive pairs and maximizing the distance between embeddings of anchor and negative pairs within each batch. The loss for a given triplet (\(a\), \(p\), \(n\)) is mathematically defined as:

\begin{equation}
L(a, p, n) = \max(0, d(E_a, E_p) - d(E_a, E_n) + \text{margin})
\end{equation}

where \(d(E_x, E_y)\) calculates the distance between two embeddings \(E_x\) and \(E_y\), and \(\text{margin}\) is a predefined margin that enforces a minimum distance between the anchor-positive and anchor-negative pairs.

The batch loss is computed as the mean of the losses for all triplets within the batch:

\begin{equation}
L_{\text{SBCL}} = \frac{1}{|\mathcal{T}|} \sum_{(a, p, n) \in \mathcal{T}} L(a, p, n)
\end{equation}

where \(\mathcal{T}\) denotes the set of all triplets in the batch. This formulation ensures the development of an embedding space that accurately represents the distinctions between security and non-security instances, facilitating effective classification.

\subsection{Prediction and Training Layer for Security Patch Detection}

The final component of \toolname is a Training and Prediction Layer, specifically designed for security patch detection. 
This layer is responsible for interpreting the fused embeddings produced by {\em PT-Former} and making accurate predictions regarding the security relevance of each patch.

\noindent
\textbf{Training Procedure.}
Training the model to accurately predict security patches involves minimizing a loss function that measures the discrepancy between the predicted probabilities and the ground-truth labels. A commonly used loss function for binary classification tasks is the binary cross-entropy loss, given by:
\begin{equation}
    L_{BCE} = -\frac{1}{N} \sum_{i=1}^{N} [y_i \log(P_i) + (1 - y_i) \log(1 - P_i)]
\end{equation}
where \(N\) is the number of examples in the training set, \(y_i\) is the ground-truth label for the \(i\)-th example (1 for security-related and 0 for non-security-related), and \(P_i\) is the computed probability for the \(i\)-th example to be security-related.

In an end-to-end training regime, both the contrastive loss from the previous sections and the BCE loss are combined: $L= L_{BCE} + L_{SBCL}$. 
At the end of the training, a learned weight matrix is available to drive inference.

\noindent
\textbf{Prediction Step.} 
The prediction mechanism utilizes a fully connected (FC) neural network layer that takes as input the fused embedding from  {\em PT-Former}, representing the unified view of the patch, its generated explanation, the developer description, and \toolname instruction. 
The FC layer is defined as follows:
\begin{equation}
    P = \sigma(W_p \cdot E + b_p)
\end{equation}
where \(E\) denotes the single fused embedding input, \(W_p\) is the learned weight matrix of the FC layer, \(b_p\) is the bias term, and \(\sigma\) represents the activation function, typically a sigmoid function for binary classification tasks such as security patch detection. The output \(P\) signifies the probability that a given patch is security-relevant.

%% file: sections/4.experiemtal_design.tex
\section{Experimental Setup}
We discuss the research questions that we are investigating, before presenting the baselines and datasets as well as the evaluation metrics.

\subsection{Research Questions}

\begin{itemize}[leftmargin=*]
    \item \textbf{RQ.1} {\em How effective is {\em \toolname} in identifying security patches?} We assess \toolname against well-known literature benchmarks and compare the achieved performance against some strong baselines.
    \item \textbf{RQ-2} {\em How do key design decisions in \toolname contribute to its performance?} We perform an ablation study where we investigate the added value of label-wise training, the generated explanations, PT-Former and contrastive learning.  
    \item \textbf{RQ-3} {\em To what extent the distribution of patch representations in \toolname improves over the state of the art?} We visualize the learned representations from  \toolname and GraphSPD to observe the differences in their potential discriminative power. Based on case studies, we also qualitatively assess how \toolname representation assigns scores to key tokens. 
    \item \textbf{RQ-4} {\em Does the trained \toolname model generalize beyond our study dataset?} We evaluate the robustness of \toolname by applying the model trained on a given dataset to samples from a different dataset. 
\end{itemize}

\subsection{Datasets}
We consider two datasets from the recent literature :
\begin{itemize}[leftmargin=*]
\item \textbf{PatchDB}~\cite{wang2021patchdb} is an extensive set of patches of C/C++ programs. It includes about 12K security-relevant and about 24K non-security-relevant patches. The dataset was constructed by considering patches referenced in the National Vulnerability Database (NVD) as well as patches extracted from GitHub commits of 311 open-source projects (e.g., Linux kernel, MySQL, OpenSSL, etc.). 
\item \textbf{SPI-DB}~\cite{zhou2021spi} is another large dataset for security patch identification.
The public version includes patches from FFmpeg and QEMU, amounting to about 25k patches (10k security-relevant and 15k non-security-relevant).  
\end{itemize}

We selected the aforementioned datasets because they collectively provide a significant variety in the vulnerabilities as well as a spectrum of patches (with different styles, syntax and semantic implementations). Thus, they are suitable for intra-project and cross-project assessment.

\subsection{Evaluation Metrics}

We consider common evaluation metrics from the literature:

\begin{itemize}[leftmargin=*]
\item {\bf +Recall} and {\bf -Recall}. These metrics are borrowed from the field of patch correctness prediction~\cite{tian2022change}. In this study, +Recall measures a model's proficiency in predicting security patches, whereas -Recall evaluates its capability to exclude non-security ones.

\item {\bf AUC and F1-score~\cite{hossin2015review}}. 
The overall effectiveness of \toolname is gauged using the AUC  (Area Under Curve) and F1-score metrics. 
\end{itemize}

\subsection{Baseline Methods}
\begin{itemize}[leftmargin=*]
\item \textbf{GraphSPD}: We consider the most recently published state-of-the-art GraphSPD~\cite{wang2023graphspd}, which, after demonstrating that prior token-based approaches do not capture sufficient semantics, deploys a cutting-edge graph neural network method for security patch detection. Indeed, it represents a significant advancement by using graph representations of patches, allowing for richer semantics compared to previous deep neural network methods relying on token sequences.

\item \textbf{TwinRNN}: In our study, we opt for RNN-based solutions~\cite{wang2021patchrnn}\cite{zhou2021spi}, which leverage a twin RNN architecture to assess the security relevance of a given patch. This approach involves employing two RNN modules, each equipped with shared weights, to analyze the code sequences before and after the patch application. 

\item \textbf{GPT}: 
We consider LLMs as relevant baseline given that we employ them as part of our pipeline (to generate patch explanations). We opt for GPT (v3.5)~\cite{brown2020language}, which is accessible.  
We prompt it with the following instructions: 
\textit{``Given the following code change, determine if it is related to a security vulnerability or not. Please respond with either `security' or `non-security' and you must provide an answer. [{\tt Patch information}]"}


\item \textbf{CodeT5}: Similarly to GPT, because the CodeT5~\cite{wang2021codet5} encoder-decoder model is a core component that is used as an initial embedder of patches in \toolname, we consider it as a baseline approach for classifying patches.

\item \textbf{VulFixMiner}~\cite{zhou2021finding} builds on the CodeBERT transformer-based approach for representing patches to train the security patch identification classifier. We reproduce it as a baseline. 

\end{itemize}

Beyond these baselines, the literature in software engineering has recently proposed  {\bf CoLeFunDa}~\cite{zhou2023colefunda}. However, we do not directly compare against it in our work because it is closed-source\footnote{We have requested access to the code. However, the authors have replied that they are not authorized to share it by their employer - Huawei.} and not readily reproducible.

With CoLeFunDa, the authors propose to use the GumTree differencing tool to extract the description of changes that are made. It considers syntactic descriptions of change operations (e.g., UPDATE invocation at IF) while our approach generates descriptions that provide step by step reasoning. It should be noted that its major benefit is visible in terms of effort-based metrics. In the original publication, the authors show that it improves over VulFixMiner by 1\% in terms of AUC.

\subsection{Implementation}
We develop \toolname using the Pytorch library (version 11) and run our experiments on two V100 (32GB) GPUs with the cuda-11 version. We take AdamW~\cite{loshchilov2018fixing} as weight optimization. 
We run a total of 20 epochs with a learning rate of 1e-05 and a decay rate of 0.01 to achieve convergence and regularization. Batch sizes of 16 for training and 64 for testing are chosen to facilitate smooth workflow. Alpha, temperature, and dropout parameters are set to 0.5, 0.1, and 0.5, respectively.

%% file: sections/5.experiment_result.tex
\section{Experiment Results}

\input{sections/Exps/RQ1}

\input{sections/Exps/RQ2}

\input{sections/Exps/RQ3}
\input{sections/Exps/RQ4}
\input{sections/Exps/RQ5}

%% file: sections/Exps/RQ1.tex
\subsection{Overall performance of \toolname}
In this section, we evaluate the performance of \toolname and compare against the selected baselines across the PatchDB and SPI-DB datasets. 
Table~\ref{tab:comparison} reports the performance measurements on different metrics. 

\begin{table}[h!]
    \centering
    \caption{Performance metrics (\%) on security patch detection}
    \resizebox{\linewidth}{!}{
    \label{tab:comparison}
    \begin{tabular}{c|c|c|c|c|c}
        \hline

        \hline

        \hline

        \hline
       \bf Method & \bf Dataset & \bf AUC & \bf F1 & \bf +Recall & \bf -Recall \\
        \midrule
        \multirow{2}{*}{$\begin{array}{c}\text{TwinRNN} \\
        \text{\cite{wang2021patchrnn}}\end{array}$} & PatchDB & 66.50 & 45.12 & 46.35 & 54.37 \\
        & SPI-DB & 55.10 & 47.25 & 48.00 & 52.10  \\
        \hline 
        \multirow{2}{*}{$\begin{array}{c}\text{GraphSPD} \\
        \text{\cite{wang2023graphspd}}\end{array}$} & PatchDB & 78.29 & 54.73 & 75.17 & 79.67\\
        & SPI-DB & 63.04 & 48.42 & 60.29 & 65.33  \\
        \hline 
        \multirow{2}{*}{$\begin{array}{c}\text{GPT (v3.5)} \\
        \end{array}$} & PatchDB & 50.01 & 52.97 & 49.28 & 50.67\\
        & SPI-DB & 49.83 & 42.19 & 44.70 & 55.20  \\
        \hline 
        \multirow{2}{*}{$\begin{array}{c}\text{Vulfixminer~\cite{zhou2021finding}} \\
        \end{array}$} & PatchDB & 71.39 & 64.55 & 55.72 & 77.03 \\
        & SPI-DB & 68.04 & 54.42 & 68.14 & 62.04 \\
        \hline 
        \multirow{2}{*}{$\begin{array}{c}\text{CodeT5~\cite{wang2021codet5}} \\
        \end{array}$} & PatchDB & 71.00 & 63.73 & 54.98 & 76.18 \\
        & SPI-DB & 72.88 & 56.77 & 65.45 & 68.75 \\
        \hline
        \multirow{2}{*}{\textbf{LLMDA}} & PatchDB & \cellcolor{black!25}84.49 ($\pm$ 0.51) & \cellcolor{black!25}78.19 ($\pm$ 0.37) & \cellcolor{black!25}80.22 ($\pm$ 0.21) & \cellcolor{black!25}87.33 ($\pm$ 0.24) \\
        & SPI-DB & \cellcolor{black!25}68.98 ($\pm$ 0.27) & \cellcolor{black!25}58.13 ($\pm$ 0.33) & \cellcolor{black!25}70.94 ($\pm$ 0.13) & \cellcolor{black!25}80.62 ($\pm 0.22$)  \\
        \hline

        \hline

        \hline

        \hline
    \end{tabular}}
\end{table}

\toolname is consistently able to identify security patches and recognize non-security patches. On the PatchDB dataset, this performance reaches 80\% and 87\%, respectively, for +Recall and -Recall. On the SPI-DB dataset, the performance is lower but, again, consistent across both classes. 

The results achieved by the baselines (cf. Table~\ref{tab:comparison}) further demonstrate the superior performance of \toolname. On the PatchDB dataset, \toolname significantly outperforms token-driven neural network approaches, including VulfixMiner, TwinRNN, and GPT 3.5 on all metrics. The performance improvement ranges from $\sim$18\% to $\sim$24\% in terms of AUC. This large improvement is also noticeable in the other metrics and with the SPI-DB dataset. 

With respect to the incumbent state-of-the-art, GraphSPD, we note that \toolname outperforms it by about 6, 23, 5, and 8 percentage points, respectively, in terms of AUC, F1, +Recall, and -Recall on the PatchDB dataset. On the SPI-DB dataset, the metric improvements are also substantial: 5 (AUC), 10 (F1), 10 (+Recall) and 15 (-Recall) percentage points.

\find{{\bf  [RQ-1] \ding{42} } \toolname is effective in detecting security patches. With an F1 score at 78.19\%, \toolname demonstrates a well-balanced performance: our model can concurrently attain high precision and high recall. Specifically, we achieved a new state-of-the-art performance in identifying both security patches (+Recall) and recognizing non-security patches (-Recall). Comparison experiments further confirm that \toolname is superior to the baselines and is consistently high-performing across the datasets and across the metrics. 
}


%% file: sections/Exps/RQ2.tex
\subsection{Contributions of key design decisions} 
In this section we investigate the impact of key design choices on the overall performance of \toolname. To that end, we perform an ablation study on :
\begin{itemize}
    \item {\em Inputs}: Compared to prior works, \toolname innovates by considering two additional inputs, namely an LLM-generated explanation of the code changes as well as an instruction. What performance gain do we achieve thanks to these inputs? 
    \item {\em Representations:} A major contribution of the \toolname design is the {\em PT-Former} module, which enables to align and concatenate bimodal input representations belonging to different embedding spaces. What performance gap is filled by PT-Former?
    \item {\em Classifiers:} \toolname relies on stochastic batch contrastive learning to enhance its discriminative power, in particular for samples that are close to the decision boundaries of security relevance. To what extent does SBCL maximize \toolname's performance?
\end{itemize}

To answer the aforementioned sub-questions, we build variants of \toolname where different components are removed. We then compute the performance metrics of each variant and compare them against the original \toolname.

\subsubsection{\bf \em Impact of LLM-generated explanations}
We build a variant {\sc llmda}$_{EX-}$ where the explanation input is replaced by ``[CLS]''. Indeed, changing the PT-Former architecture to consider three inputs may bias the experiments. Instead, we follow the convention recognized by transformer-based models: the ``[CLS]'' token represents the classification token and is positioned at the beginning of the input sequence, serving as a signal for the model to generate a representation suitable for classification tasks. In our case, since the instruction input also indicates that the representation is for a classification task, our replacement has no side effect. 

{\sc llmda}$_{EX-}$ allows us to investigate the model's performance when the contextual information provided by the LLM-generated explanation is not provided. Table~\ref{tb:woex} reports the performance results that are achieved in this ablation study.

\begin{table}[h!]
    \caption{Performance (\%) of {\sc llmda}$_{EX-}$ (without the LLM-generated explanations)}
    \centering
    \resizebox{\linewidth}{!}{
    \begin{threeparttable}  
    \begin{tabular}{c|c|c|c|c|c}
        \hline

        \hline

        \hline

        \hline 
        Model & Dataset & AUC & F1 & +Recall & -Recall  \\
        \midrule
        \multirow{2}{*}{{\sc llmda}$_{EX-}$}
        & PatchDB & 83.24 (\textcolor{red}{$\downarrow1.25$})\tnote{$*$} & 76.73 (\textcolor{red}{$\downarrow1.46$})\tnote{$*$} & 79.01 & 86.09   \\
        & SPI-DB & 68.27 (\textcolor{red}{$\downarrow0.71$})\tnote{$*$} & 57.57 (\textcolor{red}{$\downarrow0.56$})\tnote{$*$} & 70.23 & 80.07  \\
        \hline

        \hline

        \hline

        \hline
    \end{tabular}
    \begin{tablenotes}
    \item[*] \textcolor{red}{($\downarrow x.xx$)} measures the performance drop when comparing against \toolname.
  \end{tablenotes}
    \end{threeparttable}}
    \label{tb:woex}
\end{table}

 It is noticeable that the performance of {\sc llmda}$_{EX-}$ is consistently lower across the various metrics and across the datasets. These findings highlight the significance of LLM-generated explanations in enhancing the model's predictive capabilities.

\subsubsection{\bf \em Impact of instruction}
We conduct an ablation study based on a variant, {\sc llmda}$_{IN-}$, where the designed instruction is replaced with the ``[CLS]'' token as in the previous ablation study. The performance of this variant on the PatchDB and SPI-DB datasets is reported in Table~\ref{tb:woin}.

\begin{table}[h!]
    \caption{Performance (\%) of {\sc llmda}$_{IN-}$ (without the designed instruction)}
    \centering
    \resizebox{\linewidth}{!}{
    \label{tab:ablationstudy}
    \begin{tabular}{c|c|c|c|c|c}
        \hline

        \hline

        \hline

        \hline 
        Model & Dataset & AUC & F1 & +Recall & -Recall \\
        \midrule
        \multirow{2}{*}{\sc llmda$_{IN-}$} 
        & PatchDB & 82.51 (\textcolor{red}{$\downarrow 1.98$}) & 76.14 (\textcolor{red}{$\downarrow 2.05$}) & 78.55 & 85.64   \\
        & SPI-DB & 67.93 (\textcolor{red}{$\downarrow 1.05$}) & 57.25 (\textcolor{red}{$\downarrow 0.88$}) & 69.90 & 79.62  \\
        \hline

        \hline

        \hline

        \hline
    \end{tabular}}
    \label{tb:woin}
\end{table}

Again, we note that the performance drops compared to \toolname. It even appears that, without the instruction, the performance drop is slightly more important than when the model does not include the LLM-generated explanations. These findings underscore the importance of the label-wise design decision based on explicitly adding an instruction among the inputs for embedding to enhance the model's performance for security patch detection.

\subsubsection{\bf \em Impact of PT-Former}
To investigate the importance of {\em PT-Former}, we design a variant,  {\sc llmda}$_{PT-}$ where the {\em PT-Former} space alignment \& representation combination module is removed. To that end, we must still design some simple computations to generate one embedding space for all inputs. In our case, the embeddings of the code and text input tokens have the same dimension ($768$). We thus concatenate them:
\begin{equation}
E = \mathrm{Avg}(E_{p}+E_{expl}) \oplus E_{desc} \oplus E_{inst}
\end{equation}

where $\mathrm{Avg}$ is the average operation and $\oplus$ is the concatenation operation.

\begin{table}[h!]
    \caption{Performance (\%) of {\sc llmda$_{PT-}$} (without the {\em PT-Former} module)}
    \centering
    \resizebox{\linewidth}{!}{
    \label{tab:ablationstudy}
    \begin{tabular}{c|c|c|c|c|c}
        \hline

        \hline

        \hline

        \hline 
        Model & Dataset & AUC & F1 & +Recall & -Recall  \\
        \midrule
        \multirow{2}{*}{\sc llmda$_{PT-}$} 
        & PatchDB & 80.36 (\textcolor{red}{$\downarrow 4.13$}) & 70.24 (\textcolor{red}{$\downarrow 7.95$}) & 69.54 & 83.18  \\
        & SPI-DB & 63.17 (\textcolor{red}{$\downarrow 5.81$}) & 54.62 (\textcolor{red}{$\downarrow 3.51$})& 67.15 & 73.51   \\
        \hline

        \hline

        \hline

        \hline
    \end{tabular}}
    \label{tb:wopt}
    \vspace{-0.2cm}
\end{table}

The performance results of {\sc llmda$_PT-$} are reported in Table~\ref{tb:wopt}. Compared to other variants of \toolname, {\sc llmda$_{PT-}$} achieves the lowest scores across all evaluated metrics. On PatchDB, compared to the original \toolname, {\sc llmda$_{PT-}$} performance is dropped by 4.13\%, 7.95\% in terms of AUC and F1. Actually, {\sc llmda$_{PT-}$} has over 10\% less +Recall, meaning that it is significantly under-performing in the task of identifying security patches. On SPI-DB, the performance gap is larger on -Recall (it fails to recognize non-security patches). These findings confirm that the design decisions in {PT-Former} have been instrumental to the performance of \toolname.

\subsubsection{\bf \em Impact of SBCL}
\begin{figure}[ht]
    \vspace{-0.5cm}
    \centering
    \includegraphics[width=0.8\linewidth]{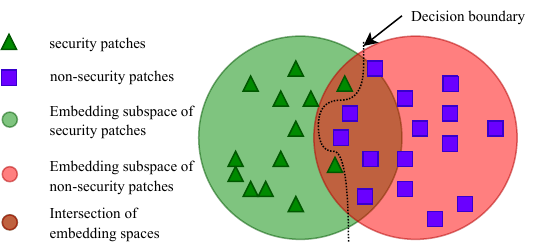}
    \caption{Illustration of embedding subspaces of security/non-security patches for contrastive learning}
    \label{fig:sbclwo}
    \vspace{-0.2cm}
\end{figure}

In \toolname we designed SBCL to optimize the model's ability to discern different patterns between positive (security patches) and negative (non-security patches) examples more effectively.  Figure~\ref{fig:sbclwo} illustrates the ambition: after PT-Former learns the representations, the embedding subspaces of security and non-security patches will certainly intersect on some ``difficult'' samples. SBCL is designed to find the optimum decision boundary. 
To assess the importance of SBCL, we design a variant, {\sc llmda$_{SBCL-}$}, where we directly feed the embeddings processed by \textit{PT-Former} into the fully connected layer (i.e., without the stochastic batch contrastive learning step).

Table~\ref{tb:wosbcl} presents the performance results of {\sc llmda$_{SBCL-}$}. Compared to the original \toolname, the performance drop is noticeable. Despite the relatively small proportion of the semantic space at the intersection between the subspaces of security and non-security patches, SBCL enables to achieve 1-3 percentage points improvement on the different metrics.

\begin{table}[h!]
    \caption{Performance (\%) of {\sc llmda$_{SBCL-}$} (without contrastive learning)}
    \centering
    \resizebox{\linewidth}{!}{
    \label{tab:ablationstudy}
    \begin{tabular}{c|c|c|c|c|c}
        \hline

        \hline

        \hline

        \hline 
        Model & Dataset & AUC & F1 & +Recall & -Recall   \\
        \midrule
        \multirow{2}{*}{\sc llmda$_{SBCL-}$} 
        & PatchDB & 82.93 (\textcolor{red}{$\downarrow 1.56$}) & 76.45 (\textcolor{red}{$\downarrow 1.74$}) & 78.72 & 85.81  \\
        & SPI-DB & 67.43 (\textcolor{red}{$\downarrow 1.55$}) & 56.61 (\textcolor{red}{$\downarrow 1.52$}) & 69.45 & 79.10 \\
        \hline

        \hline

        \hline

        \hline
    \end{tabular}}
    \label{tb:wosbcl}
\end{table}


\find{{\bf  [RQ-2] \ding{42} }
The ablation study results reveal that each of the key design decisions contributes noticeably to the performance of \toolname. In particular, without the {\em PT-Former} module \toolname would lose about 8 percentage points in F1. 
}

%% file: sections/Exps/RQ3.tex
\subsection{Discriminative power of \toolname representations}\label{sec:type}
In this section, we investigate to what extent the representations obtained with \toolname are indeed enabling a good separation of security and non-security patches in the embedding space. To that end, we consider two separate evaluations: the first attempts to visualize the embedding space of \toolname and compares it against the one of GraphSPD (i.e., the state of the art); the second qualitatively assesses two case studies.

\subsubsection{\bf \em Visualization of embedding spaces}
\begin{figure}[!h]
    \centering
    \begin{subfigure}[b]{0.4\textwidth}
    \centering
        \includegraphics[width=0.75\columnwidth]{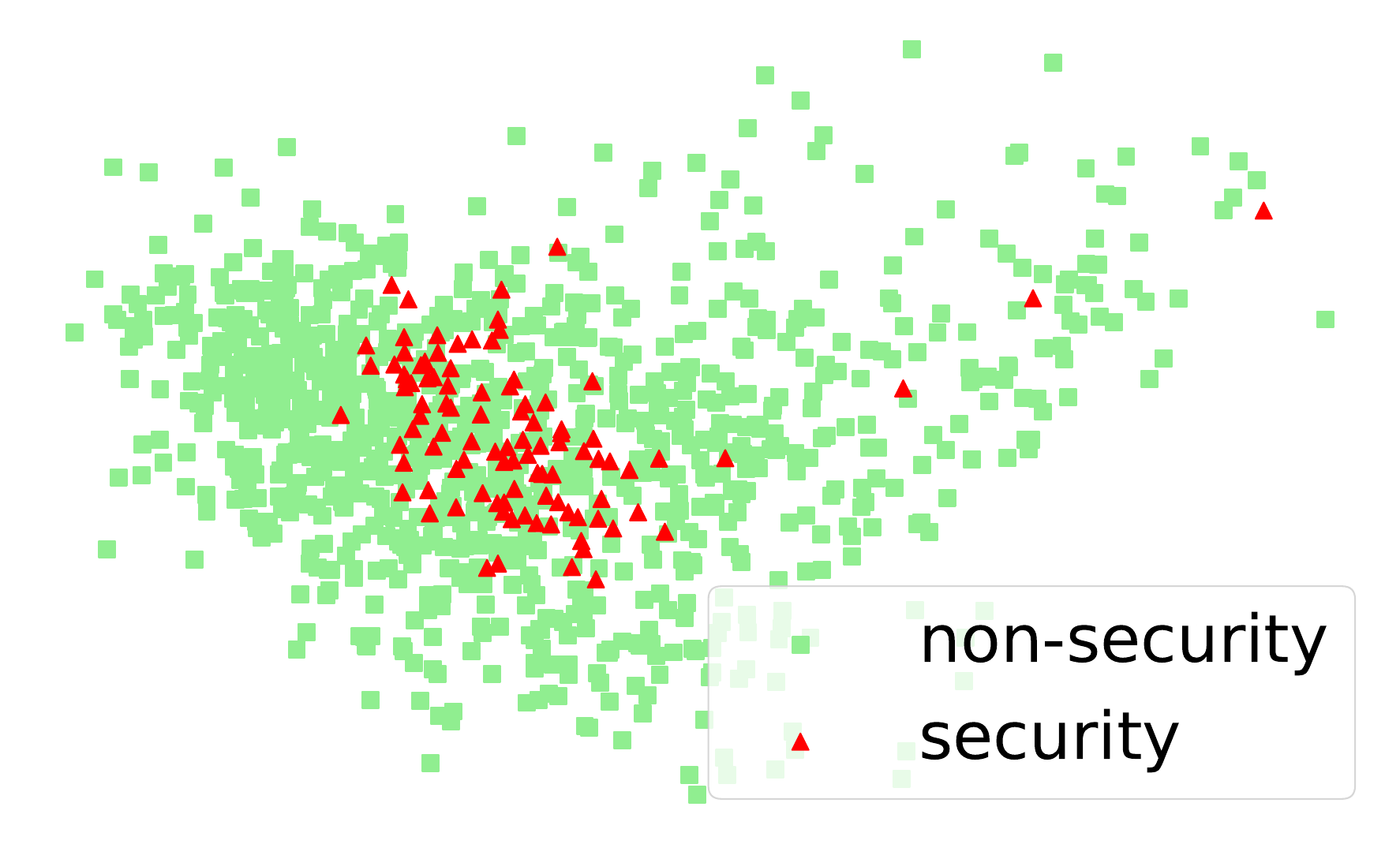}
        \caption{Embeddings yielded by GraphSPD}
        \label{fig:sub1}
    \end{subfigure}
    \begin{subfigure}[b]{0.4\textwidth}
        \includegraphics[width=0.75\columnwidth]{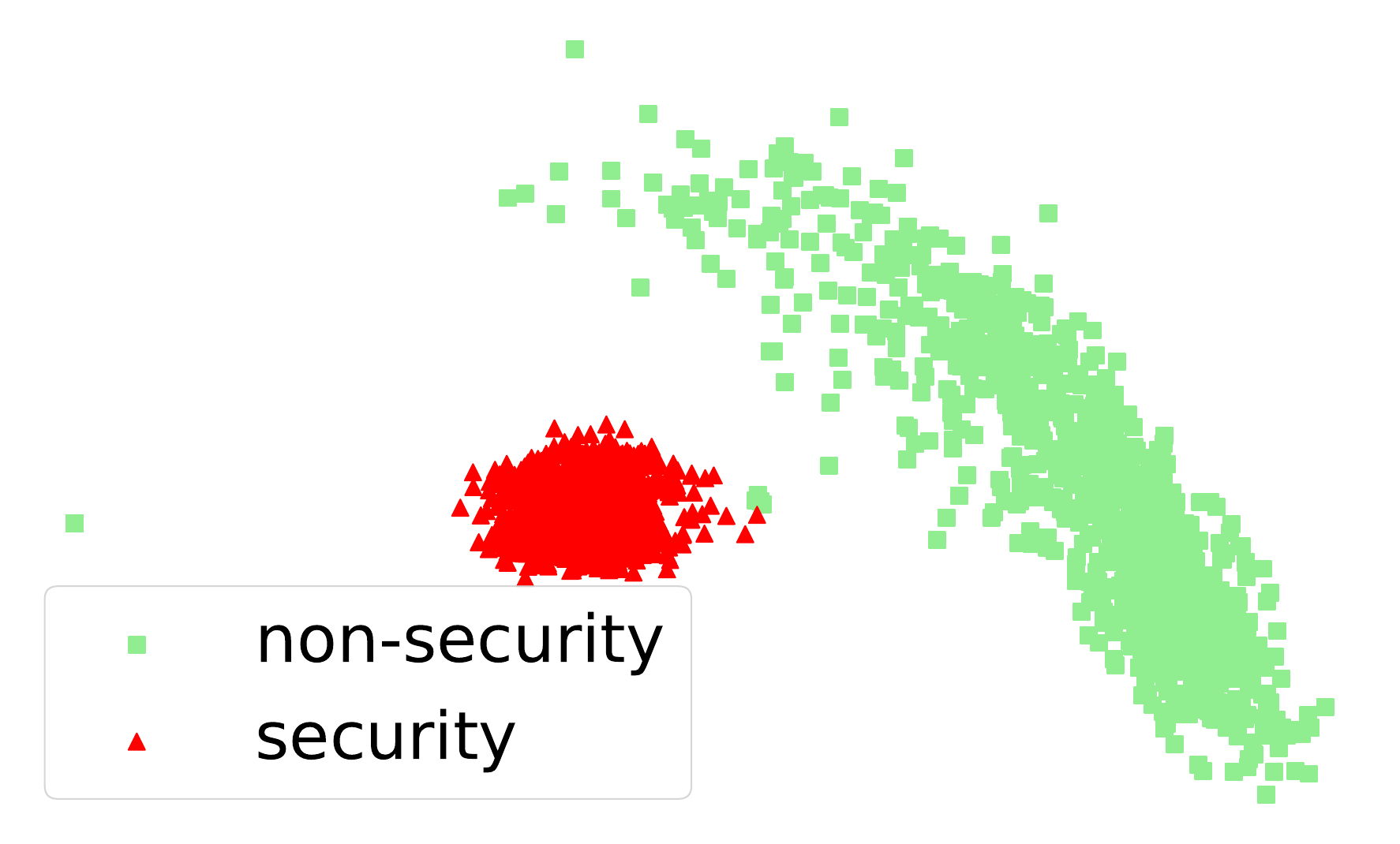}
        \caption{Embeddings yielded by \toolname}
        \label{fig:sub2}
    \end{subfigure}
    \caption{PCA visualizations of security and non-security patch embeddings by GraphsSPDand \toolname.
    }
    \label{fig:visual}
\end{figure}

We consider 1\,000 random patches from our PatchDB dataset. We then collect their associated embeddings from \toolname and GraphSPD and apply principal component analysis (PCA)~\cite{smith2002tutorial}. Given the imbalance of the dataset, the drawn samples are largely non-security patches, while security patches are fewer. Figure~\ref{fig:visual} presents the PCA visualizations of the representations.

\begin{table*}[!t]
    \centering
    \caption{Attention scores for security and non-security labels by GraphSPD and \toolname on two sample patches}
    \resizebox{1\linewidth}{!}{
\begin{tabular}{>{\centering}m{0.4\linewidth}|m{0.2\linewidth}|>{\centering}m{0.4\linewidth}|m{0.2\linewidth}|m{0.2\linewidth}}
\hline

\hline

\hline

\hline

\multicolumn{2}{c|}{GraphSPD (failed cases)} & \multicolumn{3}{c}{\toolname (successful cases)} \\
    \cline{1-5}
 Patch ({\em non-formatted token sequence}) & Developer Description& Patch ({\em non-formatted token sequence})& Explanation & Description \\ \hline
 
\scriptsize \tt diff --git a/sgminer.c b/sgminer.cindex a7dd3ab3..08697cd0 100644--- a/sgminer.c+++ b/sgminer.c@@ -518,7 +518,7 @@ struct pool \textcolor{red}{*add\_pool(void)}\textcolor{blue}{ \rm \em  \textbf{(non-security score = 0.65)}}sprintf(buf, "Pool \%d", pool\texttt{->}\_no);pool\texttt{->}poolname = strdup(buf);-\textbackslash tpools = realloc(pools, sizeof(struct pool *) \*(total\_pools + 2));+\textbackslash tpools = (struct **pool **)realloc(pools, sizeof(struct pool *) * (total\_pools + 2));pools[total\_pools++] = pool;mutex\_init(\&pool\texttt{->}pool\_lock); if(unlikely({pthread\_cond\_init}(\&pool\texttt{->}cr\_cond, NULL))) 
& Fixed missing realloc removed by mistake & \scriptsize \tt diff --git a/sgminer.c b/sgminer.cindex a7dd3ab3..08697cd0 100644--- a/sgminer.c+++ b/sgminer.c@@ -518,7 +518,7 @@ struct pool *add\_pool(void)sprintf(buf, "Pool \%d", pool\texttt{->}\_no);pool\texttt{->}poolname = strdup(buf);-\textbackslash tpools = \textcolor{red}{\textbf{realloc(pools, sizeof(struct pool *) \*(total\_pools + 2));+\textbackslash tpools = (struct **pool **)realloc(pools, sizeof(struct pool *) * (total\_pools + 2));}} \textcolor{blue}{\rm \em \textbf{(security core = 0.57)}} pools[total\_pools++] = pool;\textcolor{red}{\textbf{mutex\_init}}\textcolor{blue}{\rm \em \textbf{(security core = 0.50)}}(\&pool\texttt{->}pool\_lock); if(unlikely({pthread\_cond\_init}(\&pool\texttt{->}cr\_cond, NULL)))  & Modified sgminer.c, adjusted \textcolor{red}{\textbf{realloc usage}}\textcolor{blue}{\rm \em \textbf{(security core = 0.67)}} with \textcolor{red}{\textbf{explicit type casting}}\textcolor{blue}{\rm \em \textbf{(security core = 0.63)}} & Fixed missing \textcolor{red}{\textbf{realloc}}\textcolor{blue}{\rm \em 
 \textbf{(security core = 0.63)}} removed by mistake\\ 
\hline

\hline

\hline

\hline

\scriptsize \tt diff --git a/lib/krb5/auth\_context.c b/lib/krb5/auth\_context.c index 0edea5418..3cba484e1 100644 --- a/lib/krb5/auth\_context.c +++ b/lib/krb5/auth\_context.c @@ -53,6 +53,7 @@ \textcolor{red}{\textbf{krb5\_auth\_con\_init}}\textcolor{blue}{\rm \em \textbf{(non-security score = 0.60)}}(krb5\_context context, \textcolor{red}{\textbf{ALLOC}}\textcolor{blue}{\rm \em 
 \textbf{(non-security score = 0.67)}}(p\texttt{->}authenticator, 1); if (!p\texttt{->}authenticator) return ENOMEM; +    memset (p\texttt{->}authenticator, 0, sizeof(*p\texttt{->}authenticator)); p\texttt{->}flags = KRB5\_AUTH\_CONTEXT\_DO\_TIME; 
& zero authenticator
& \scriptsize \tt diff --git a/lib/krb5/auth\_context.c b/lib/krb5/auth\_context.c index 0edea5418..3cba484e1 100644 --- a/lib/krb5/auth\_context.c +++ b/lib/krb5/auth\_context.c @@ -53,6 +53,7 @@ krb5\_auth\_con\_init(krb5\_context context, ALLOC(p\texttt{->}authenticator, 1); if (!p\texttt{->}authenticator) return ENOMEM; +    \textcolor{red}{\textbf{memset (p\texttt{->}authenticator, 0, sizeof(*p\texttt{->}authenticator));}}\textcolor{blue}{\rm \em \textbf{(security core = 0.59)}} p\texttt{->}flags = KRB5\_AUTH\_CONTEXT\_DO\_TIME; 
& Added
memset to initialize \textcolor{red}{\textbf{`authenticator' memory in krb5\_auth\_
con\_init function}}\textcolor{blue}{\rm \em \textbf{(security core = 0.84)}}. & \textcolor{red}{\textbf{zero authenticator}} \textcolor{blue}{\rm \em \textbf{(security core = 0.77)}}\\
\hline

\hline

\hline

\hline
\end{tabular}}
    \label{tb:case}
\end{table*}

 We observe from the distribution of data points that  \toolname can effectively separate the two categories (i.e., security and non-security patches), in contrast to the incumbent state-of-the-art, GraphSPD. This finding suggests that the representations of \toolname are highly relevant for the task of security patch detection. 

%% file: sections/Exps/RQ4.tex
\subsubsection{\bf \em Case studies}
Table~\ref{tb:case} presents 2 examples to illustrate the difference between \toolname and GraphSPD in terms of what the representations can capture, and potentially explaining why \toolname was successful on these cases while GraphSPD was not. For our classification task, we have two labels: security (0) and non-security (1). 
For \toolname, we can directly consider the label name in the instruction. Thus we compute the attention map between security and the tokens in the patch, the explanation, and the description. For  GraphSPD, however, since there no real label name involved in the training and inference phases, we compute the attention score between the words in the patch and the number ``0" or ``1". 
To simplify the analysis, we only highlight, in Table~\ref{tb:case}, tokens for which the similarity score is higher than 0.5.

As shown in the examples, \toolname generally assigns high similarity scores to security-related aspects, suggesting a detection capability that nuances between tokens. For example, in the sgminer.c patch, \toolname gives high scores to \textit{realloc} and \textit{mutex\_init}, indicating a finer sensitivity to potential security implications within these code parts. Similarly, in the krb5/auth\_context.c patch, the use of \textit{memset} for initializing authenticator memory is scored high in \toolname, reflecting its more acute recognition of security practices.

In contrast, since GraphSPD is graph-based, it focuses on the patch itself. For the same patch cases, GraphSPD can even give  very high attention scores for non-security label. For example, \textit{krb5\_auth\_con\_init} is given 0.6 score for ``non-security" and \textit{ALLOC} is given 0.67 attention score towards non-security as well. These scores may justify many failures of GraphSPD in the security patch detection task.

\find{{\bf  [RQ-3] \ding{42} }
The design of \toolname leads to patch representations that enable enhanced ability over GraphSPD in effectively differentiating between security and non-security patches on the embedding spaces. 
Our analysis of sample cases shows that \toolname assigns high attention scores to tokens associated to security-related aspects, making it effective for accurately identifying security patches.}




%% file: sections/Exps/RQ5.tex
\subsection{Robustness of \toolname} 
A model is accepted as robust if it performs strongly on datasets that differ from the training data. For our study task, robustness should ensure reliable predictions on unseen patches. We assess the  
robustness of \toolname and GraphSPD by training them against the PatchDB and testing against the samples from the FFmpeg dataset used to construct the benchmark for Devign~\cite{zhou2019devign} vulnerability detector. This test data includes 13,962 data points, consisting of 8,000 security-related and 5,962 non-security-related parches. The selection of the FFmpeg dataset is motivated by its coverage of a wide range of vulnerabilities.

\begin{table}[htbp]
    \centering
    \caption{Performance (\%) of GraphSPD and \toolname against unseen patches from the FFmpeg dataset~\cite{zhou2019devign}. \newline {\scriptsize Numbers between parentheses ($\downarrow$X) corresponds to the drop of performance when compared to the evaluation in cross-validation with PatchDB}}
    \resizebox{\linewidth}{!}{
    \label{tb:robustness}
    \begin{threeparttable}
 
    \begin{tabular}{c|c|c|c|c|c|c|c}
        \hline

        \hline

        \hline

        \hline
        Method & Accuracy & Precision & AUC & Recall & +Recall & -Recall & F1 \\
        \midrule
        GraphSPD & 43.65 & 51.15 & 44.81 & 36.88 & 36.88 (\textcolor{red}{$\mathbf{\downarrow}$\bf 38.29}) & 52.75 & 42.86 (\textcolor{red}{$\downarrow$\bf 11.74)}\\ \hline
        \toolname & \cellcolor{black!25}66.78 & \cellcolor{black!25}72.70 & \cellcolor{black!25}66.69 & \cellcolor{black!25}67.30 & \cellcolor{black!25}67.30 (\textcolor{red}{$\downarrow$12.92}) & \cellcolor{black!25}66.09 & \cellcolor{black!25}69.89 (\textcolor{red}{$\downarrow$8.3})\\
        \hline

        \hline

        \hline

        \hline
    \end{tabular}
         \begin{tablenotes}
    \item[*] (\textcolor{red}{$\downarrow x.xx$}) measures the performance drop when comparing with the cross-validation of \toolname on Patch-DB (cf. Table~\ref{tab:comparison}).
  \end{tablenotes}      
    \end{threeparttable}
    }
\end{table}

Table~\ref{tb:robustness} summarizes the performance results of  \toolname and GraphSPD on the unseen dataset. Overall, on all metrics, \toolname exhibits a significantly superior performance  over GraphSPD, with about 20 percentage points of gap in terms of precision for example. We further highlight in the results the performance decrease between the test on unseen data and the cross-validation test in terms of F1 and +Recall (i.e., the ability to identify security patches). We note that the robustness of \toolname is substantially higher than GraphSPD:  GraphSPD loses about 38 percentage points of +Recall when \toolname only loses about 13 points.

\find{{\bf  [RQ-4] \ding{42} }
Experiments on unseen patches clearly demonstrate that \toolname is more robust than GraphSPD. In terms of the ability to identify security patches, GraphSPD performance is dropped about threefold compared to \toolname under the same experimental settings.}

%% file: sections/8.threats.tex
\section{Discussion}
\subsection{Threats to Validity}
\noindent {\bf Internal validity.} A first threat is the quality of the generated patch explanations. Since LLMs may be factually wrong in their descriptions of the code changes or, in contrast, be vastly good for our well-known study datasets, \toolname performance evaluation may be biased. We mitigate this threat by considering a state-of-the-art LLM and by rigorously analyzing the impact of the generated LLM in an ablation study.

A second threat is the evolving performance of the hosted GPT models. It may prevent reproducibility since this evolution introduces instability, potentially affecting the consistency of results even with identical prompts or instructions. 

A third threat lies in the constraint imposed by the input size limitation to 512 tokens. For long patches, \toolname performs truncation, resulting in the loss of essential information and potentially affecting the accuracy and reliability of the model's predictions. 

\noindent {\bf External validity.} A threat is that we rely on PatchDB and SPI-DB datasets, which may not generalize our findings beyond their diverse samples. For example, SPI-DB contains patches from only 2 projects. We mitigate this threat by relying on 2 distinct datasets, PatchDB having samples from over 300 projects. Furthermore \toolname is natural language-centric and thus our key design choices are programming language-independent.

Another threat stems from the fact that we rely on pre-trained models (CodeT5 and LlaMa-7b) as initial embedders of \toolname's inputs. These models may actually not be adapted for the task at hand. To mitigate this threat our selection was based on the fact that they were demonstrated in the literature as among the best performing models for related tasks.

\noindent {\bf Construct validity.} A threat is that our experiments do not try various prompts in the {\em Instruction} input. This may lead to an oversight in properly checking the potential contribution of instruction on the model's performance. We have mitigated this threat by performing an ablation study that shows the impact of the current design. Better prompts may positively increase this impact. 

\subsection{Limitation}
One limitation of the current study is the use of GPT-3.5 for generating explanations of code changes within the \toolname framework. While GPT-3.5 has demonstrated strong performance in understanding and explaining code, the release of GPT-4.0 presents an opportunity for further enhancement.

To illustrate this, consider the following patch (non-security) from the Linux kernel, which modifies socket configurations to enable asynchronous I/O operations:
\begin{lstlisting}[language=diff]
@@ -1950,6 +1950,7 @@ static int xs_local_finish_connecting(struct rpc_xprt *xprt,
        sk->sk_user_data = xprt;
        sk->sk_data_ready = xs_data_ready;
        sk->sk_write_space = xs_udp_write_space;
+		sock_set_flag(sk, SOCK_FASYNC);
        sk->sk_error_report = xs_error_report;
        sk->sk_allocation = GFP_NOIO;

@@ -2136,6 +2137,7 @@ static void xs_udp_finish_connecting(struct rpc_xprt *xprt, struct socket *sock)
        sk->sk_user_data = xprt;
        sk->sk_data_ready = xs_data_ready;
        sk->sk_write_space = xs_udp_write_space;
+		sock_set_flag(sk, SOCK_FASYNC);
        sk->sk_allocation = GFP_NOIO;

        xprt_set_connected(xprt);

@@ -2237,6 +2239,7 @@ static int xs_tcp_finish_connecting(struct rpc_xprt *xprt, struct socket *sock)
        sk->sk_data_ready = xs_tcp_data_ready;
        sk->sk_state_change = xs_tcp_state_change;
        sk->sk_write_space = xs_tcp_write_space;
+		sock_set_flag(sk, SOCK_FASYNC);
        sk->sk_error_report = xs_error_report;
        sk->sk_allocation = GFP_NOIO;
\end{lstlisting}

GPT-4.0 provides a more detailed and contextually rich explanation:
\begin{quote}
The provided code snippets modify socket configurations in three different functions (xs\_local\_finish\_connecting, xs\_udp\_finish\_connecting, xs\_tcp\_finish\_connecting). They uniformly set the SOCK\_FASYNC flag using \texttt{sock\_set\_flag(sk, SOCK\_FASYNC);}, enabling asynchronous I/O operations on the sockets involved. This change likely aims to enhance efficiency by allowing non-blocking handling of network events and data reception across different types of connections (local, UDP, TCP).
\end{quote}

In contrast, GPT-3.5's explanation is more concise:
\begin{quote}
The changes in the code snippets add the \texttt{sock\_set\_flag(sk, SOCK\_FASYNC);} line to three different functions (xs\_local\_finish\_connecting, xs\_udp\_finish\_connecting, and xs\_tcp\_finish\_connecting). This modification enables asynchronous (signal-driven) I/O on the sockets, allowing the program to receive signals when I/O events occur, thereby improving network operations' efficiency and responsiveness.
\end{quote}

While both explanations are accurate, GPT-4.0's explanation offers a more nuanced understanding of the code changes' implications, which could be crucial for determining the security relevance of a patch. In this case, the patch is likely non-security-related, as it focuses on improving I/O efficiency rather than addressing a security vulnerability.

Future work could explore the integration of GPT-4.0 or other advanced language models to enhance the \toolname framework's ability to generate detailed explanations and improve the prediction of security patches. This could lead to more accurate and timely detection of silent security patches, thereby enhancing the overall security posture of open-source software systems.

%% file: sections/2.relatedwork.tex
\section{Related Work}\label{sec:relatedwork}
Our work is related to various research directions in the literature. We discuss three main categories in this section.
\subsection{Security Patch Analysis.}
Patch analysis, after being addressed in the literature of empirical studies and static analysis research, has been increasingly a key application area of machine learning for software engineering~\cite{li2017large,wang2023graphspd,wang2020machine,perl2015vccfinder}. In terms of security patches, Li et al.~\cite{li2017large} provided foundational empirical insights into the unique attributes of such  patches. 
Rule-based approaches~\cite{wu2020precisely,huang2019using} were then pivotal in demonstrating that the identification of security patches is feasible using common patterns~\cite{xu2020automatic,xu2020automatic}. Afterwards, Wang et al.~\cite{wang2020machine} proceeded to data-driven methodologies with statistical machine learning. RNN-based approaches such as PatchRNN~\cite{wang2021patchrnn} and SPI~\cite{zhou2021spi} then revealed that neural networks were key enablers in understanding patches. With ColeFunda~\cite{zhou2023colefunda}, researchers proposed to summarize the semantics of patches using git differencing tools. Most recently, GraphSPD~\cite{wang2023graphspd} achieved state-of-the-art performance by implementing a graph-based approach that focuses on ensuring that the semantics in the code change are effectively captured.
In this work, our \toolname approach employs Large Language Models for semantic analysis of code changes and introduces a multi-modal alignment method to improve the accuracy of security patch detection.

\subsection{Deep Learning in Vulnerability Detection.}
Deep learning has enabled software engineering research to advance in the automation of the detection of vulnerable code~\cite{li2021sysevr,fu2024aibughunter,russell2018automated}. 
Most recently, Fu et al. advanced software vulnerability detection by proposing VulExplainer~\cite{fu2023vulexplainer} for the classification of vulnerability types using Transformer-based hierarchical distillation. In another direction, Nguyen et al. contributed by identifying vulnerability-relevant code statements through deep learning and clustered contrastive learning~\cite{nguyen2022information} and by creating ReGVD~\cite{nguyen2022regvd}, a graph neural network model for vulnerability detection.

\subsection{Patch representation learning} 
Reasoning about patches using deep neural networks has attracted significant interest in recent years. While initial works directly leveraged generic code representation models such as CodeBERT~\cite{feng2020codebert}, CodeT5~\cite{wang2021codet5}, GraphCodeBERT~\cite{guo2020graphcodebert} or PLBART~\cite{ahmad2021unified}. Some recent works, such as CCRep~\cite{liu2023ccrep}, ReconPatch~\cite{hyun2024reconpatch}, CCBERT~\cite{zhou2023ccbert}  have explored specialized approaches to better capture semantics of code changes. With \toolname, our approach attempts to learn specific representations for the task of security patch detection. Our approach, \toolname, builds on the foundation of leveraging deep neural networks for patch representation, advancing beyond generic models like CodeBERT and CodeT5 by focusing on specialized representation learning tailored specifically for detecting security patches.

%% file: sections/6.conclusion.tex
\section{Conclusion}
\label{sec:conclusion}
In this work, we proposed a framework, \toolname, for security patch detection. It implements a language-centric approach to the overall problem of learning to identify silent security patches. First, \toolname augments patch information with LLM-generated explanations. Then, it builds an embedding where multi-modal patch information are concatenated with an natural language instruction after the alignment of embedding spaces. Finally, using contrastive learning, it ensures that challenging cases are the decision boundaries are well discriminated. Experimental assessment of \toolname over two literature datasets demonstrate how \toolname achieves new state of the performance on the target task. Further ablation studies confirm the contribution of the key design choices as well as the robustness of the trained model.

\noindent {\bf Open Science:} All code, data and results are publicly available in our artefacts repository: \textcolor{blue}{\url{https://llmda.github.io}}